\documentclass[aps,preprint,groupedaddress,showkeys]{revtex4}
\usepackage{graphicx}
\usepackage{setspace}

\begin{document}

\singlespacing

 \title{DNA sequence correlations shape nonspecific transcription factor-DNA binding affinity} 
\author {Itamar Sela and David B. Lukatsky$^{*}$}
\affiliation{Department of Chemistry, Ben-Gurion University of the Negev, 84105 Beer-Sheva, Israel}

\begin{abstract}
Transcription factors (TFs) are regulatory proteins that  bind DNA in promoter regions of the genome and
either promote or repress gene expression.
Here we predict analytically that enhanced homo-oligonucleotide sequence correlations, such as poly(dA:dT) and
poly(dC:dG) tracts, statistically
enhance non-specific TF-DNA binding affinity. This prediction is generic and qualitatively independent of
microscopic parameters of the model. We show that non-specific TF binding affinity is universally controlled 
by the strength and symmetry of DNA sequence correlations. We perform correlation analysis of 
the yeast genome and show that DNA regions highly occupied by TFs exhibit stronger homo-oligonucleotide sequence correlations,
and thus higher propensity for non-specific binding,
as compared with poorly occupied regions. We suggest that this effect plays the role of an effective localization potential
enhancing the quasi-one-dimensional diffusion of TFs in the vicinity of DNA, speeding up the stochastic search process
for specific TF binding sites. The predicted effect also imposes an upper bound on the size of TF-DNA binding motifs.
\end{abstract}

\keywords{Promiscuity of transcription factor-DNA binding; Free energy of transcription factor-DNA binding}

\maketitle

\section{Introduction}

Transcription factors (TFs) are proteins that regulate gene expression in both prokaryotic ({\it{e.g.}} bacteria) and eukaryotic ({\it{e.g.}} yeast or human)
cells. TFs bind regulatory promoter regions of DNA in the genome. It is commonly accepted that each transcription factor 
binds specifically a relatively small set of DNA sequences called TF binding motifs or TF binding sites (TFBSs). A TF binds its specific 
binding motifs with a higher affinity than other genomic sequences of the same length \cite{berg87, stromo98}. A typical length of TF binding motif varies between 6 and 20 nucleotides. Recent high-throughput measurements of TF binding preferences on a genome-wide scale have challenged the classical picture of TF specificity \cite{polly,bulyk}. These experiments measured binding preferences of more than a hundred transcription factors to tens of thousands of DNA sequences and demonstrated a
high level of multi-specificity in TF binding \cite{polly,bulyk}. It has been also pointed out that weak-affinity TF binding motifs
are essential for gene expression regulation \cite{segal08}.

A key question is how TFs  find their specific binding sites in a background of $10^6-10^9$ non-specific sites in a cell genome. This question was first addressed theoretically in seminal works of Berg, Winter, and von Hippel \cite{berg81, vonhippel89}. 
The central idea of this approach is that the search process is a combination of three-dimensional and one-dimensional diffusion (see \cite{tolya11, marko04, mirny_review09} for recent reviews). It has been shown in different theoretical models that one-dimensional diffusion (in different models termed `sliding' or `hopping') facilitates the search process under certain conditions \cite{mirny04,mirny04a,mirny10, tolya08, tolya10, grosberg06,kafri09}. 
Despite the success of these phenomenological models, a complete understanding of the search process phenomena is still lacking \cite{tolya11}.
In particular, one of the key, open questions is what makes a TF switch from three-dimensional diffusion to one-dimensional sliding in specific genomic locations \cite{tolya11}.
Invariably, an assumption is made about the existence of some non-specific binding sites that bring TFs to the vicinity of DNA for one-dimensional sliding. This assumption is a key component of all theoretical models, yet the molecular origin of this effect is not understood \cite{tolya11, mirny_review09}. Recent single-molecule experimental studies undoubtedly show that different DNA-binding proteins spend the majority of their time non-specifically bound and diffusing along DNA \cite{cox06, xie09, xie07, mirny08, zhuang08}. The question is what biophysical mechanism provides such non-specific attraction towards genomic DNA and regulates the strength of this attraction at a given genomic location?  

\begin{figure}[h]
\begin{center} 
\includegraphics[width=15cm]{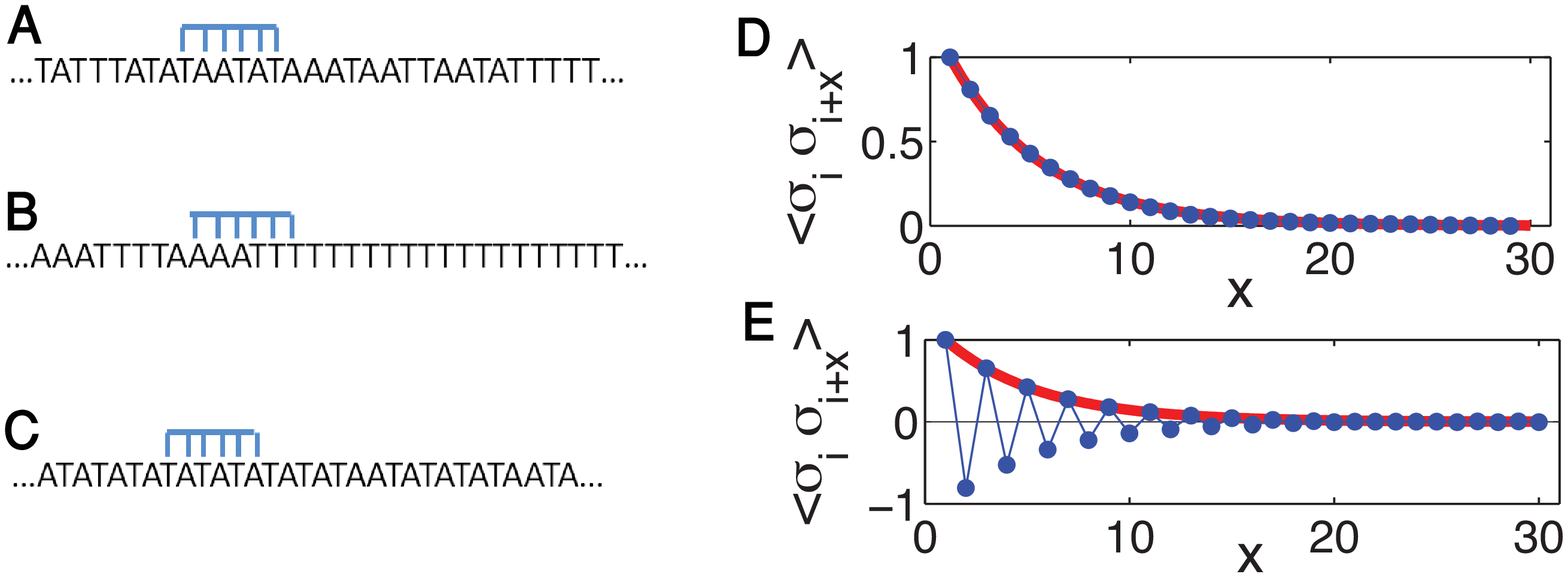}
\end{center} 
\caption{\label{fig_cartoon}}
\end{figure}

Here we predict that DNA sequence correlations statistically regulate non-specific TF-DNA binding preferences.
Depending on the symmetry and length-scale of sequence correlations, the non-specific binding affinity can be either enhanced or reduced. 
In particular, we show that homo-oligonucleotide sequence correlations, where nucleotides of the same type are clustered together 
generically reduce the non-specific TF-DNA binding free energy thus enhancing the binding affinity, Fig. \ref{fig_cartoon}. 
Sequence correlations where nucleotides of different types are alternating, lead to an opposite effect, increasing the non-specific TF-DNA binding free energy, Fig. \ref{fig_cartoon}.
Correlation analysis of the yeast genome regulatory sequences suggests that the predicted design principle is exploited at the genome-wide level, in order to increase the strength of non-specific binding at these regulatory genomic locations. 

The paper is organized as follows. First, we present a simple, analytically solvable model describing TF-DNA binding. This model
uses two-nucleotide alphabet DNA sequences. We develop a stochastic procedure 
allowing us to `design' DNA sequences with a controlled symmetry and strength of sequence correlations. We analyze the free energy
of non-specific TF-DNA binding within the framework of this model, and give an intuitive explanation for the origin of the predicted effect. 
Second, we generalise the model to four-letter alphabet
DNA sequences and show that all key conclusions hold qualitatively true in this case, as well. Third, we compute 
the free energy of non-specific TF-DNA binding for yeast genomic sequences, and show that sequences highly
occupied by TFs {\it{in vivo}}, possess a statistically higher propensity for non-specific binding to TFs, compared with sequences depleted in TFs. 
Finally, we conclude and propose experiments allowing a direct test of the predicted effect.

\section{Theory and results}

\subsection{Free energy of non-specific TF-DNA binding in model sequences}

In this work we use a simple variant of the Berg-von Hippel model to describe TF-DNA binding \cite{berg87}. For the analytical analysis we apply the model to artificial DNA sequences containing two types of nucleotides only, rather than four. However, we show that all key conclusions hold qualitatively true for four-nucleotide alphabet sequences, as well. 

The energy of a TF bound to DNA at a specific location $i$ (see Fig. \ref{fig_cartoon}):
\begin{eqnarray}
U(i) = -K\sum_{j=i}^{{\cal M}+i-1}\sigma_j, 
\label{eq_tf_en}
\end{eqnarray} 
where $i$ and $j$ represent individual base-pairs, $\cal M$ is the effective length of the TF ({\it{i.e.}} the number of contacts between TF and DNA),
$\sigma_j=\pm1$ describes two possible nucleotide types at each position $j$, and $K$ is the interaction strength.
We therefore assume that the energy contributions of individual base-pairs to the total binding energy, $U(i)$, are additive. We also assume
that the energy of each contact is exclusively defined by the base-pair type. The sequence of a DNA molecule of length $L$ is uniquely defined by the set of $L$ numbers, $\sigma_j$, with $j=1...L$. 

We note that Eq. (\ref{eq_tf_en}) provides a minimal model for TF-DNA binding. It captures the
recognition specificity of TF in a simplest possible way, by assigning different contact energies, $+K$ and $-K$, with
two possible nucleotide types. In reality, a TF recognizes DNA motifs forming a complex, cooperative network of hydrogen and electrostatic bonds \cite{berg87,stromo98}. Yet we suggest that the design principle 
for enhanced non-specific TF-DNA binding predicted using such a simplified model, is likely to be quite general
and robust with respect to microscopic details of TF-DNA interactions.

The free energy of binding of an individual TF to DNA is given by ${\cal F}=- k_B T\ln Z$, with the partition function:
\begin{eqnarray}
Z=\sum_{i=1}^L \exp(-U(i)/ k_B T), 
\label{eq_Z_single}
\end{eqnarray}
where $k_B$ is the Boltzmann constant, $T$ is the absolute temperature, and we imply periodic boundary conditions.  
We ask the question, what are the statistical properties of ${\cal F}$ as a function of the symmetry and strength of DNA sequence correlations?

In order to answer this question, we first `design' DNA sequence using a stochastic design procedure. This procedure allows nucleotides within DNA sequence to anneal, with each configuration being accepted with the Boltzmann probability: 
\begin{eqnarray}
p(E_d)=\frac{1}{Z_d} e^{-E_d/k_BT_d},
\label{eq_prob_des}
\end{eqnarray} 
where $T_d$ is the `design' temperature controlling the strength of correlations (this is different from the thermodynamic temperature, $T$), $E_d$ is the `design', intra-DNA energy. For simplicity, we take into account only the nearest-neighbor interactions in the `design' energy:  
\begin{eqnarray}
E_d = -J\sum_{i=1}^{L}\sigma_i\sigma_{i+1},
\label{eq_e_des}
\end{eqnarray} 
 with $J$ being the `design', intra-sequence interaction strength, and $Z_d$ is the corresponding Ising model partition  
 function \cite{plischke}:
 \begin{eqnarray}
Z_d &=& 2^L\left(\cosh^L\left(\beta_dJ\right) + \sinh^L\left(\beta_dJ\right)\right).
\label{eq_Zd}
 \end{eqnarray}
where $\beta_d=1/k_BT_d$. 

The ferromagnetic-like case, $J>0$, produces sequences with
homo-oligonucleotide stretches. The correlation length, $\xi=-1/ \ln(\tanh\beta_d |J|)$, 
is the characteristic length-scale of the correlations decay, $\left< \sigma_i\sigma_{i+x}\right> = \exp(-x/\xi)$ \cite{plischke}. The anti-ferromagnetic-like case, $J<0$, 
produces sequences with a different symmetry of alternating nucleotides, Fig. \ref{fig_cartoon}. 
We define the average free energy of TF binding to DNA as the annealed average: 
\begin{eqnarray}
\left<{\cal F} \right>= -\frac{1}{\beta} \ln \left<Z\right>,
\label{eq_av_F}
 \end{eqnarray}
where the averaging is performed with the probability, $p(E_d)$, Eq. (\ref{eq_prob_des}), and  $\beta=1/k_BT$. 
The quenched averaging, $\left< {\cal F}\right>_q =- \left< \ln Z \right>/\beta $, is analyzed numerically below, and
it gives qualitatively similar results, Fig. \ref{fig_Delta_F}. The averaging in Eq. (\ref{eq_av_F}) gives:
 \begin{eqnarray}
\left< Z \right> = \frac{2^{L-M-1}L}{Z_d}\Bigg [\left(\lambda_+^{\cal M} + \lambda_-^{\cal M}\right)
\left(\cosh^{L-{\cal M}}(\beta_dJ) + \sinh^{L-{\cal M}}(\beta_dJ)\right)\nonumber\\
+\left(\lambda_+^{\cal M} - \lambda_-^{\cal M}\right)
\left(\cosh^{L-{\cal M}}(\beta_dJ) - \sinh^{L-{\cal M}}(\beta_dJ)\right)\nonumber\\
\left.\times\frac{e^{-\beta_dJ}}{\sqrt{e^{2\beta_dJ}\sinh^2(\beta K)+e^{-2\beta_dJ}}}
\right],
\label{eq_av_Z}
 \end{eqnarray}
where $Z_d$ is given by Eq. (\ref{eq_Zd}), and 
 \begin{eqnarray}
\lambda_\pm &=& e^{\beta_dJ}\cosh(\beta K)\pm\sqrt{e^{2\beta_DJ}\sinh^2(\beta K)+e^{-2\beta_dJ}}.
\label{eq_lambda}
 \end{eqnarray}
 
 We argue that the DNA correlations symmetry affects statistically the interaction free energy. It is natural therefore
 to analyze the free energy difference, between `designed' sequences and their randomized analogs, lacking any symmetry:
 \begin{eqnarray}
\left<\Delta {\cal F}\right>=\left<{\cal F}\right>-\left<F_\infty\right>,
\label{eq_Delta_F}
 \end{eqnarray}
where $\left<F_\infty\right>$ is the free energy computed for entirely random sequences ({\it{i.e.}} for sequences designed using a very high value of $T_d$, or equivalently, $1/\beta_dJ \gg1$). The first key property of  $\left<\Delta {\cal F}\right>$, is that it is invariant with respect to the sign of  the TF-DNA binding affinity constant, $K$. Second, it is always satisfied that $\left<\Delta {\cal F}\right><0$ if $J>0$ (ferromagnetic-like correlations within designed DNA sequences, see Eq. (\ref{eq_e_des})), and $\left<\Delta {\cal F}\right>>0$ if $J<0$ (anti-ferromagnetic-like correlations). 
Fig. \ref{fig_Delta_F} shows the behavior of  $\left<\Delta {\cal F}\right>$ at different magnitudes of the design strength. The central observation here is that the behavior of $\left<\Delta {\cal{F}}\right>$ critically depends on the symmetry and the length-scale of DNA sequence correlations. The presence of homo-oligonucleotide stretches along DNA sequences statistically increases the propensity of such sequences towards non-specific binding to TFs. The DNA stretches with alternating nucleotides of different types produce the opposite effect: such sequences will have a reduced propensity for non-specific binding. We note that the quenched average,  $\left<\Delta {\cal{F}}\right>_q=-\left<\ln (Z/Z_\infty)\right>/\beta$, computed numerically, is in good agreement with the annealed average, Fig. \ref{fig_Delta_F}.

The reduction of the TF-DNA binding free energy by the presence of homo-oligonucleotide sequence correlations can be understood intuitively in the following way. Homo-oligonucleotide sequence correlations generically
enhance fluctuations of the TF-DNA binding energy, $\sigma_U^2=\left<U^2\right>-\left<U\right>^2$. This effect has
to do with the symmetry: a TF sliding along correlated DNA sequences where nucleotides of the same type have
the tendency to cluster, will experience homogeneous DNA `islands', such as poly(dA:dT) and poly(dC:dG) tracts. Statistically, this leads to the dominant contribution of either very strong or very weak energies to the TF-DNA binding energy spectrum. This symmetry effect leads therefore to the widening
of the TF-DNA binding energy spectrum, $P(U)$. Such widening generically leads to the reduction of the TF-DNA
binding free energy, due to the fact that the dominant contribution to the partition function, $Z$,
comes from the low-energy tail of $P(U)$ \cite{misha}. Alternatively, DNA sequence with enhanced
antiferromagnetic-like correlations ({\it{i.e.}} with alternating nucleotides of different types) will lead to the opposite effect: a TF sliding along such sequence will experience very heterogeneous binding sites. This leads
to the narrowing down of $P(U)$, and consequently, to the increase of the non-specific TF-DNA binding free energy. 

We note that the predicted effect is not restricted to TFs, and it is operational for any other kind of DNA-binding proteins.

\subsection{Extension of the model to four-letter-alphabet DNA sequences}

In the following, we show that four-letter-alphabet DNA sequences
demonstrate qualitatively
similar statistical binding properties, as those of two-letter-alphabet sequences analyzed above.
This will allow us in the following to extend all our insights gained from the analytical model directly to genomic DNA sequences.  
We argue that the same underlying physical mechanism controls the non-specific binding propensity in both cases.
 
Contrary to the two-letter-alphabet DNA sequences, where within our modeling framework a TF is fully described by the single parameter $K$, in the four-letter-alphabet DNA case, a TF is characterized by four energy parameters, $K_A$, $K_T$, $K_C$, and $K_G$. Although those energy constants are generally unknown, their order of magnitude can be roughly estimated 
as $1k_BT$, and in addition, we allow the TF-DNA contact energies to fluctuate.
We therefore draw these energies from the Gaussian probability distributions, $P(K_\alpha)$, with zero mean and standard deviations, $\sigma_\alpha$, where $\alpha=\mbox{A,T,C,G}$; and we average the free energy over many TF's realizations. 

The binding energy of TF at a given site $i$:
\begin{eqnarray}
U(i) = -\sum_{j=i}^{{\cal M}+i-1} \sum_{\alpha=1}^4 K_\alpha \sigma_j^\alpha , 
\label{eq_tf_en4}
\end{eqnarray} 
where $\sigma_j^\alpha$ is a four-component vector of the type $(\delta_{\alpha A},\delta_{\alpha T},\delta_{\alpha C},\delta_{\alpha G})$, at each DNA position $j$, with the
position of $1$ specifying one of four possible identities, (A,T,C,G), of the base-pair at the position $j$, with $\delta_{\alpha\beta}$ being the Kronecker delta.
The sequence design procedure is analogous to the one introduced above, Eq. (\ref{eq_e_des}), with the $4\times4$ symmetric matrix of the design potentials entering the sum, $-J_{\alpha\beta}\sigma_i^\alpha\sigma_{i+1}^\beta$.
The results for the average TF-DNA binding free energy in the ensemble of different TFs is shown in Fig. \ref{fig_delta_f4}. The key conclusion here is that the lower the design temperature, $T_d$, provided that in the design procedure nucleotides of the same type attract (and thus the longer the correlation length of homo-oligonucleotide stretches), the lower the TF-DNA binding free energy.

\subsection{Free energy of non-specific TF-DNA binding in yeast genome}

We ask further the key question: Is the predicted design principle for non-specific TF-DNA binding operating
in a living cell? To answer this question, we computed TF-DNA binding free energies using yeast genome
DNA. Our working hypothesis here is that if the predicted effect is operational, 
genomic regions that need to be highly-accessible by TFs should possess a higher propensity for non-specific TF-DNA binding than regions that need not be highly-accessible by TFs. To test this hypothesis we compiled two datasets of genomic DNA. First, we collected $\sim\!1600$ high-confidence yeast DNA regulatory promoter sequences (for organelle organization and biogenesis genes), each sequence $100$ nucleotide long. We use the term `upstream' to describe this dataset. These upstream sequences are experimentally known to be highly-accessible by TFs. The second dataset involves a comparable number
of weakly-accessible genomic sequences. For this purpose, we chose the first $100$ nucleotide stretches of the mRNA coding regions of those organelle organization and biogenesis genes. We use the term `downstream' to describe the second dataset. The datasets were compiled from Ref. \cite{lee07}.

It turns out that upstream sequences demonstrate statistically stronger homo-oligonucleotide correlations in A and T compared to downstream sequences, and the difference in correlations of C and G is not significant between the datasets. The normalized correlation function, $C_{AA}(x)$, computed for the sets of upstream and downstream sequences, respectively, is shown in Fig. \ref{fig_C_AA_f4}.
This function is defined as: $C_{\alpha\alpha}(x)=s_{\alpha\alpha}(x)/\left<s_{\alpha\alpha}^r(x)\right>$, where $s_{\alpha\alpha}(x)=\left<\sigma_\alpha(i)\sigma_\alpha(i+x)\right>$, and $\left<s_{\alpha\alpha}^r(x)\right>$ is obtained analogously, using the
set of randomly permuted sequences averaged with respect to different random realizations. $C_{TT}(x)$ shows
qualitatively similar behavior (data not shown).  

We now compare the TF-DNA binding free energies
for those two datasets. In order to get rid of the compositional bias, for a given TF interacting with a given DNA sequence, we always compare the difference $\Delta {\cal F}$ between the actual free energy, ${\cal F}$, and the free energy
computed for the randomized sequence (preserving the nucleotide composition of each sequence), averaged over several random realizations, $\cal F_\infty$:  $\Delta {\cal F}={\cal F}- {\cal F}_\infty$. We therefore compute 
numerically the probability distribution, $P(\Delta {\cal F})$, for these two datasets of sequences, interacting with a
model set of TFs. The TF-DNA binding contact energies, $K_\alpha$, are drawn from the Gaussian distributions, $P(K_\alpha)$, as described above. 
We stress that the only external parameters entering the model are the standard deviations, $\sigma_\alpha$, 
of $P(K_\alpha)$. In our calculations we set $\sigma_\alpha=2k_BT$ for all $\alpha$.
The computed $P(\Delta {\cal F})$ for upstream and downstream DNA sequences are shown in Fig. \ref{fig_delta_f_yeast}A. We also show the cumulative probability, at different values of the selectivity cut-off, Fig. \ref{fig_delta_f_yeast}B. 
The central conclusion here is that due to the presence of enhanced homo-oligonucleotide ({\it{i.e.}} ferromagnetic-like) sequence correlations, non-specific TF-DNA binding is statistically enhanced. At the maximal
selectivity cut-off, $\Delta {\cal F}_c\simeq -0.1 \, k_BT$ per base-pair, the probability of TF binding with the free energy below $\Delta {\cal F}_c$ to upstream
DNA regions is over $30\%$ higher than to downstream regions, Fig. \ref{fig_delta_f_yeast}B. This effect leads to the
shift of the thermodynamic equilibrium towards enhanced occupancy of TFs binding upstream regions rather than downstream regions. The average strength of the effect on TF occupancy can be estimated from the difference of the average TF-DNA
binding free energies, $\left<\Delta\Delta {\cal F}\right>=\left< \Delta{\cal F}^{up}\right>-\left< \Delta{\cal F}^{down}\right>\simeq -0.1\, k_B T$ per base-pair, 
between upstream and downstream DNA regions, respectively (difference between the peak positions in Fig. \ref{fig_delta_f_yeast}A). For a TF forming ${\cal M}$ contacts within the TF-DNA binding site, this difference 
will produce $n_{up}/n_{down}\simeq\exp(0.1\cdot{\cal M})$ shift in the relative binding occupancy, where $n_{up}$ and $n_{down}$ is the number of bound TFs in the upstream and downstream regions, respectively. For a typical TF 
forming contacts with $10$ DNA base-pairs, this leads to  $n_{up}/n_{down}\simeq 2.7$. We emphasize that the latter estimate provides 
only a lower-bound limit for the strength of the predicted correlational effect. We suggest therefore that the predicted mechanism for enhanced non-specific TF-DNA binding is operational in promoter regions of a significant fraction of yeast genes.

Finally, we note that our findings suggest the existence of an upper bound for the TF-DNA binding motif size, imposed by the maximal possible
strength of non-specific binding. It is predicted \cite{mirny04} that if the free energy of TF-DNA non-specific binding falls below $- 2\, k_B T$, this significantly slows down the sliding diffusion of TF along DNA. Our estimates therefore suggest that
such slowing down is likely when the binding motif approaches the size of $20$ base-pairs.

\section{Discussion and conclusion}

Here we predicted a generic biophysical mechanism, statistically regulating the strength of non-specific TF-DNA
binding in a genome. We showed analytically and numerically, using both artificially designed and genomic DNA sequences,
that homo-oligonucleotide correlations statistically enhance non-specific TF-DNA binding affinity. 
We used the term `ferromagnetic', to describe the symmetry of such correlations.
Alternatively, DNA sequences possessing enhanced correlations of alternating nucleotides of different types (we termed such
correlations as `anti-ferromagnetic') have a reduced propensity for non-specific binding to TFs. 

Our model description of TF-DNA binding is highly simplified. 
Yet we suggest that the design principle for enhanced
non-specific TF-DNA binding predicted in this work is likely to be quite general, it is operational in genomic locations highly occupied by TFs, and it is likely to be the rule rather than the exception. 
The robustness of our conclusions with respect to the details of the model stems from the fact that
the predicted effect arises exclusively due to DNA sequence symmetry and its strength (which is determined by the length-scale
of the correlations decay). Computational analysis of the TF-DNA binding free energy in $\sim 1600$ yeast genomic DNA regions highly
occupied by TFs shows that those regions possess much higher propensity for non-specific binding compared with
regions depleted in TFs. In our analysis we used a simple procedure to get rid of the DNA compositional bias, 
allowing us to fairly compare the relative free energies of non-specific binding in different genomic locations. 

We estimated that in yeast, the predicted effect leads to at least $\sim 0.1\, k_BT\simeq 60\, \mbox{cal/mol}$ free energy
reduction (on average) per DNA base-pair in contact with a TF, for DNA regions with enhanced
propensity for non-specific binding. This leads to at least three fold concentration enrichment in TFs (on average) of such
highly promiscuous DNA regions in yeast. 
We suggest therefore that in addition to all known signals, genomic DNA might also encode its intrinsic propensity for 
non-specific binding to TFs. The predicted effect plays the role of an effective, non-specific localization potential, enhancing
the level of one-dimensional  diffusion of TFs along genomic DNA at the genome-wide level, and thus speeding up the search process for specific TF
binding sites \cite{berg81,vonhippel89,tolya11,marko04,mirny_review09,mirny04}. We stress that all our conclusions are obtained assuming a quasi-equilibrium nature of TF-DNA binding. It would be important 
to investigate the dynamic aspects of the predicted phenomena.

It is important to note that too high level of non-specific TF-DNA binding impairs the overall search efficiency \cite{mirny04,mirny04a}. This suggests that the strength of the predicted effect  {\it{in vivo}} might be subject to both positive and negative regulation. It has been pointed out in a seminal work of Iyer and Struhl \cite{struhl95} that 
activity of poly(dA:dT) tracts increases with their length. We suggest that this observation is a direct consequence
of the effect of enhanced non-specific TF-DNA binding by poly(dA:dT), predicted here. Another key observation
of Iyer and Struhl \cite{struhl95}, that poly(dC:dG) functions in a similar manner to poly(dA:dT), further strengthens our prediction.

Extensive correlation analysis of different organismal genomes and direct, large-scale measurements of TF-DNA 
binding preferences using DNA sequences with the controlled strength and symmetry of correlations, should provide
an ultimate test of the phenomenon predicted here. Protein-DNA binding arrays \cite{bulyk} and high-throughput microfluidics
technology \cite{polly} allow a direct experimental test of our predictions {\it{in vitro}}. A key experiment would
measure the TF-DNA binding affinity in different sets of DNA, each set containing DNA sequences with a specific TF-DNA binding motif 
embedded in a background of non-specific sequences with a varying symmetry and strength of correlations between DNA sets. 
We expect that DNA sequences with enhanced homo-oligonucleotide correlations in background sequences, will generically
possess a higher binding affinity to different TFs compared with background sequences either lacking such correlations or 
having correlations with alternating nucleotides of different types. 

\begin{acknowledgments}
We thank Noa Musa-Lempel for help in compiling yeast genomic sequences. 
D. B. L. acknowledges the financial support from the Israel Science Foundation grant 1014/09. 
\end{acknowledgments}

$^*$Corresponding author. Email: lukatsky@bgu.ac.il

%\newpage

\newpage
%FIGURES
\textbf{FIGURE LEGENDS}

\begin{figure}[h]
\begin{center} 
\includegraphics[width=15cm]{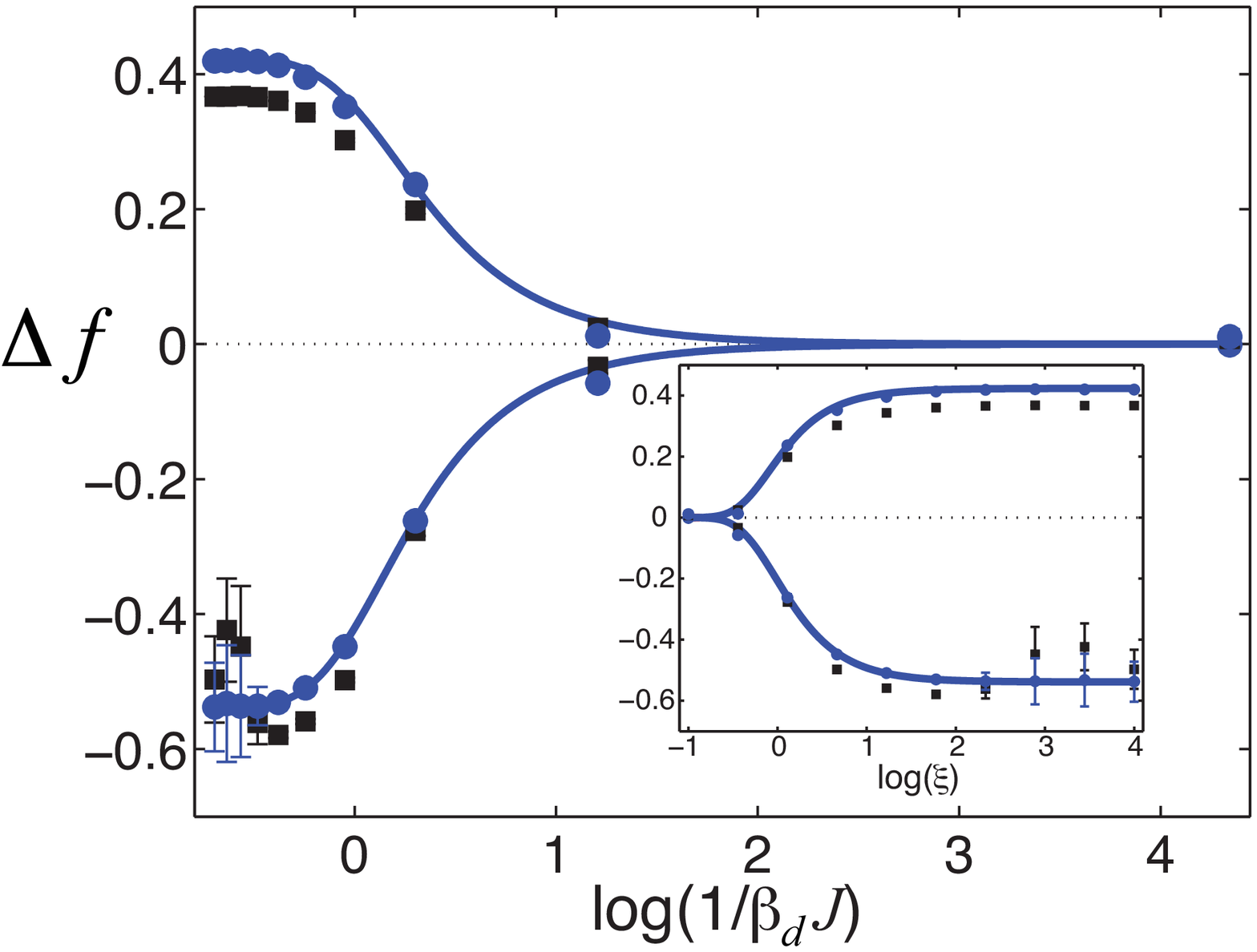}
\end{center} 
\caption{  \label{fig_Delta_F}}
\end{figure}

\begin{figure}[h]
\begin{center} 
\includegraphics[width=15cm]{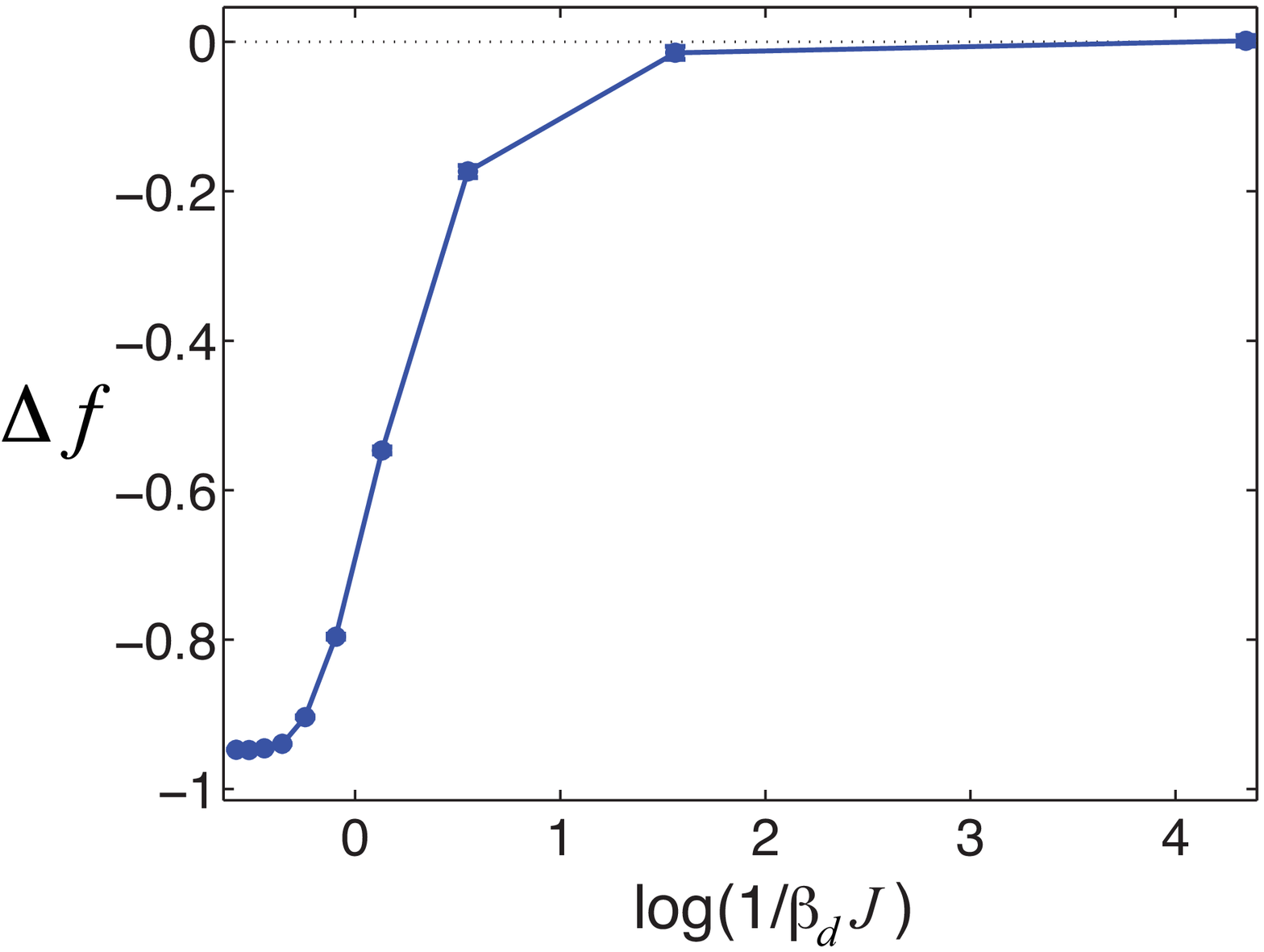}
\end{center} 
\caption{\label{fig_delta_f4}}
\end{figure}

\begin{figure}[h]
\begin{center} 
\includegraphics[width=15cm]{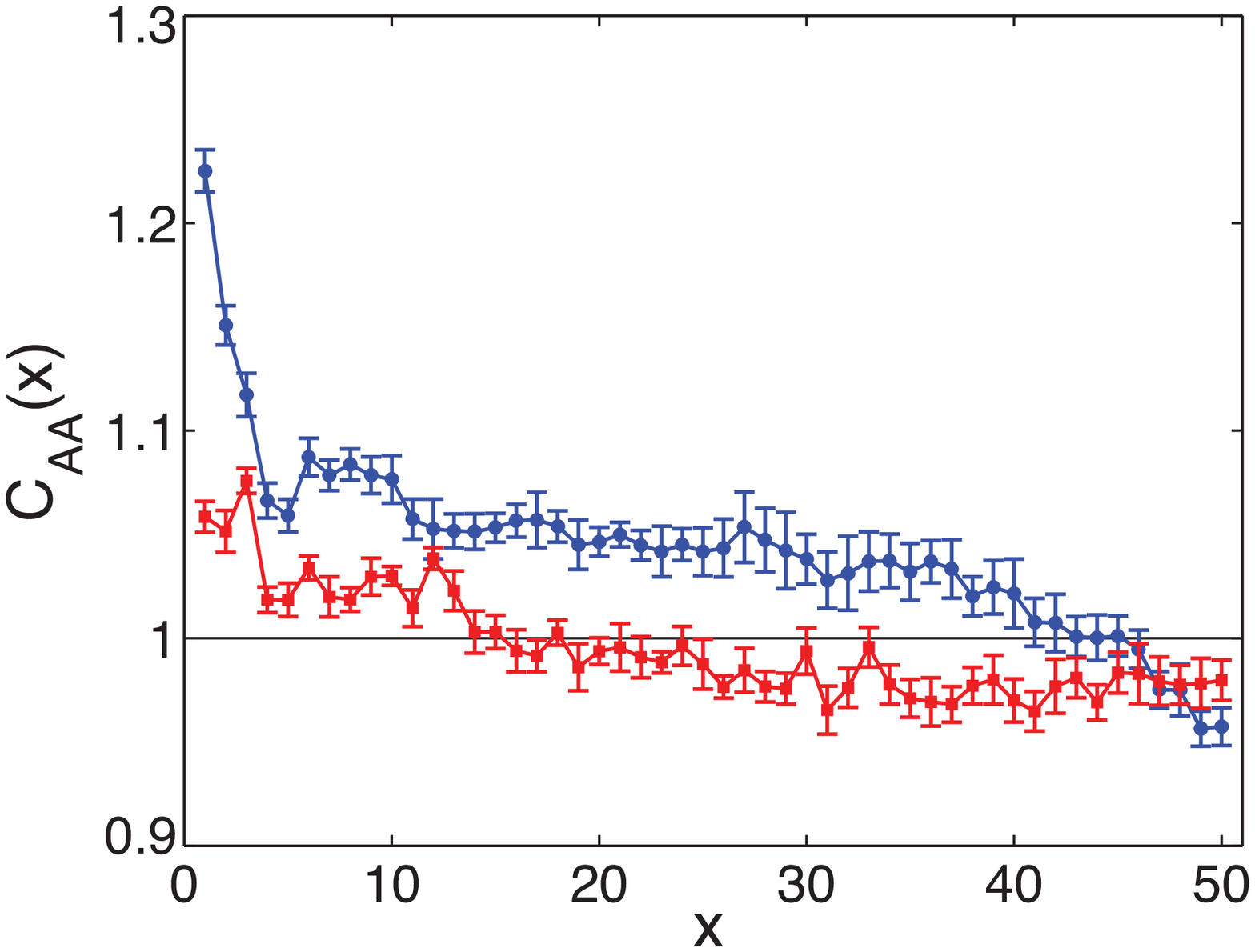}
\end{center} 
\caption{ \label{fig_C_AA_f4}}
\end{figure}

\begin{figure}[h]
\begin{center} 
\includegraphics[width=15cm]{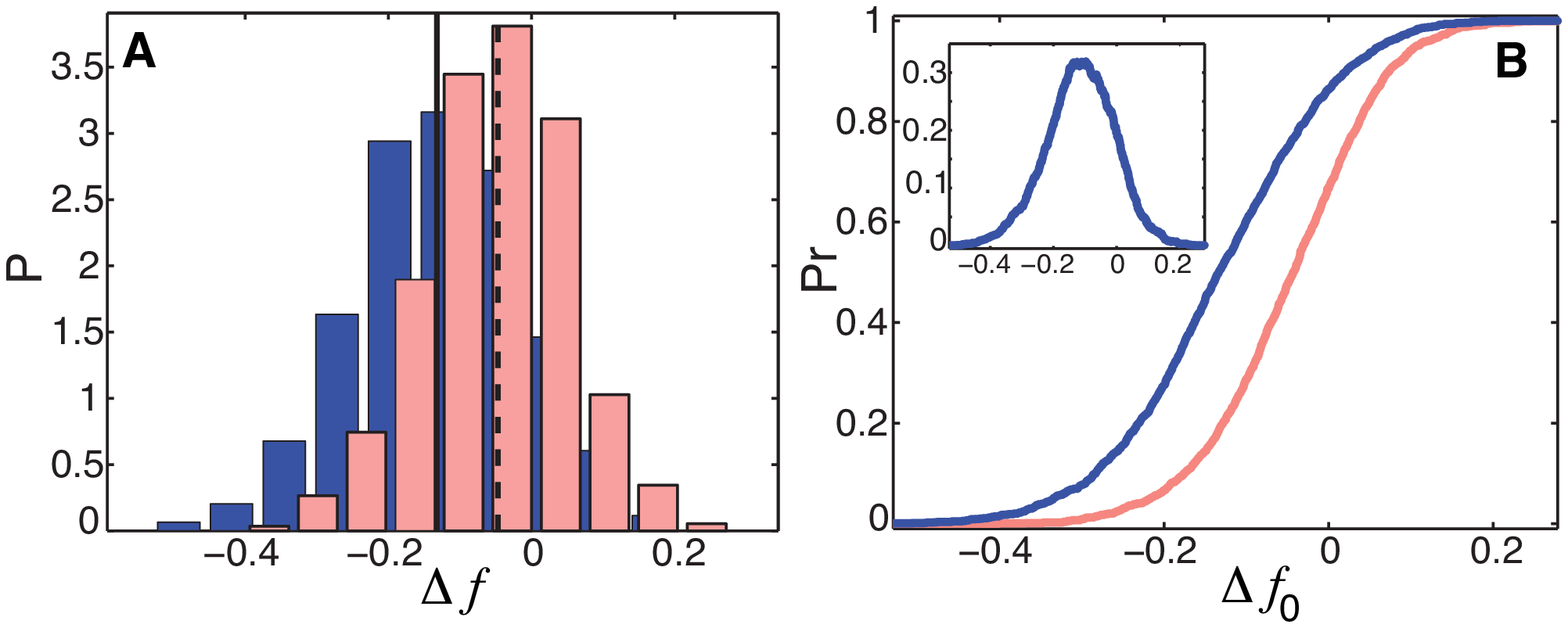}
\end{center} 
\caption{\label{fig_delta_f_yeast}}
\end{figure}

\begin{description}
  \item[FIG. 1:] Schematic representation of the model for TF binding to DNA, and examples of DNA sequence correlation functions.
{\bf{A.}} Random sequence. {\bf{B.}} Enhanced homo-oligonucleotide ({\it{i.e.}} ferromagnetic-like) correlations lead to statistically enhanced non-specific TF-DNA binding affinity.
{\bf{C.}} Enhanced anti-ferromagnetic-like correlations (alternating nucleotides of different types) lead to reduced non-specific TF-DNA binding affinity.
All examples of sequences represent simulation snapshots. 
{\bf{D.}} Example of the correlation function computed for sequences with enhanced ferromagnetic-like correlations; and 
{\bf{E.}} for sequences with enhanced anti-ferromagnetic-like correlations. The bold lines represent the exponential decay of the correlation functions.

  \item[FIG. 2:] TF-DNA binding free energy difference normalized per one base-pair, $\Delta f=\beta\left<\Delta {\cal F}\right>/{\cal M}$, computed using Eq. (\ref{eq_av_Z}) as a function of the reduced design temperature, $1/\beta_dJ$ (solid curves). The upper and lower branches of the graph correspond to $J<0$ (anti-ferromagnetic-like DNA sequence correlations) and $J>0$ (ferromagnetic-like correlations), respectively. The results of MC simulations
of the system are in excellent agreement with  the analytical results (filled circles). We used the parameters:
$\beta K=1 $, ${\cal M}=18$, $L=1000$. In MC simulations we used $7.5 \times 10^6$ MC moves to design each DNA sequence at each value of $T_d$. In order to generate each point in the plot we used the set of $100$ sequences. In order to compute error bars we divided each set of $100$ sequences into $10$ subsets randomly,
and then calculated standard deviation of the subsets averages for $\Delta f$. The error bars correspond to one standard deviation. The numerically computed quenched average, $-\left< \ln (Z/Z_\infty)\right>/{\cal M}$, is also shown (filled squares). In the computations we used the same parameters and definitions as specified above. Inset: The same data for $\Delta f$ as in the main figure, but plotted as a function of $\xi$.

  \item[FIG. 3:] The average TF-DNA binding free energy, $\Delta f$, numerically computed at different values of the design temperature, where $\Delta f=-\left< \ln (Z/Z_\infty)\right>/{\cal M}$, where $Z_\infty$ is the partition function for entirely random DNA sequence. We designed $200$ sequences with  the length $L=400$ at each $T_d$. We performed $5\times 10^{6}$ MC steps to design each sequence, where in each MC step we attempted to exchange two base-pairs chosen at random. The overall nucleotide composition for each sequence was uniform and fixed. The design potential was $+J$ (attraction) for identical nearest-neighbor base-pairs and $-J$ (repulsion) for different nearest-neighbor base-pairs, with $J=1k_BT$. The contact energies, $K_\alpha$, were drawn from a Gaussian distribution, $P(K_\alpha)$, with zero mean, $\left<K_\alpha\right>=0$, and standard deviation, $\sigma_\alpha=2k_BT$, for each nucleotide type, $\alpha$.
We  computed $\Delta f$ as an average over $250$ TFs and $200$ sequences at each $T_d$, and used ${{\cal M}=8}$. The error bars are calculated as specified in Fig. \ref{fig_Delta_F}, and they are smaller than the marker size.

  \item[FIG. 4:] The normalized correlation function, $C_{AA}(x)$ (see the text for the definition), computed for upstream (circles) 
and downstream (squares) sequence sets. Each set consists of  $1,663$ sequences; each sequence is $100$ nucleotide long.

  \item[FIG. 5:]{\bf A.} The computed  $P(\Delta f)$ for $1,663$ upstream (dark) and downstream (bright) 
yeast genomic sequences, where $\Delta f=\beta\left<\Delta {\cal F}\right>_{TF}/{\cal M}$, and $\Delta {\cal F}={\cal F}- {\cal F}_\infty$.  For each given TF, ${\cal F}_\infty$ is computed as an average over $50$ randomized sequence replicas (randomization preserves the nucleotide composition of each sequence). For each sequence we computed $\Delta {\cal F}$ for $250$ TFs and then took the average
of these $250$ values, $\left<\Delta {\cal F}\right>_{TF}$. We used ${\cal M}=8$. The TF-DNA contact energies, $K_\alpha$,  are drawn from a Gaussian probability distribution, $P(K_\alpha)$, with zero mean and standard deviation $\sigma_\alpha=2k_BT$, where $\alpha$ represents four possible nucleotides. Vertical lines show the mean of $\Delta f$.
{\bf B.} The cumulative probability, $Pr(\Delta f\leq\Delta f_0)=\int_0^{\Delta f_0} P(\Delta f) d \Delta f$,
computed using $P(\Delta f)$ from ({\bf A}), for upstream (dark) and downstream (bright) sequences, respectively. Inset: The difference between upstream and downstream $Pr(\Delta f\leq\Delta f_0)$. 
\end{description}

\newpage

%\newpage

\begin{thebibliography}{1}

\bibitem{berg87}
Berg, O.~G., and P.~H. von Hippel. 1987. 
Selection of DNA binding sites by regulatory proteins. Statistical-mechanical theory and application to operators and promoters. 
{\it{J. Mol. Biol.}} {\bf{193}}:723-750. 

\bibitem{stromo98}
Stromo,G.~D., and D.~S. Fields. 1998. 
Specificity, free energy and information content in proteinÐDNA interactions. 
{\it{Trends Biochem. Sci.}} {\bf{23}}:109-113.

\bibitem{polly} 
Fordyce, P.~M., D. Gerber, D. Tran, J. Zheng, ..., S.~R. Quake. 2010.
De novo identification and biophysical characterization of transcription-factor binding sites with microfluidic affinity analysis.
{\it{Nature Biotech.}} {\bf{28}}:970-975.

\bibitem{bulyk}
Badis, G., M.~ F. Berger, A.~A. Philippakis, S. Talukder, ..., M.~L. Bulyk. 2009.
Diversity and Complexity in DNA Recognition by Transcription Factors.
{\it{Science}}. {\bf{324}}:1720-1723.

\bibitem{segal08}
Segal, E., T. Raveh-Sadka, M. Schroeder, U. Unnerstall, and U. Gaul. 2008.
Predicting expression patterns from regulatory sequence in Drosophila segmentation.
{\it{Nature}}. {\bf{451}}:535:540.

\bibitem{berg81}
Berg, O.~G., R.~B. Winter, and P.~H. von Hippel. 1981.  
Diffusion-driven mechanisms of protein translocation on nucleic acids. 1. Models and theory.
{\it{J. Biol. Chem.}} {\bf{20}}:6929-6948.

\bibitem{vonhippel89}
von Hippel, P.~H., and O.~G. Berg. 1989.
Facilitated target location in biological systems.
{\it{J. Biol. Chem.}} {\bf{264}}:675-678.

\bibitem{tolya11}
Kolomeisky, A.~B. 2011.
Physics of protein-DNA interactions: mechanisms of facilitated target search.
{\it{Phys. Chem. Chem. Phys.}} {\bf{13}}:2088-2095.

\bibitem{marko04}
Halford, S.~E., and J. F. Marko. 2004.
How do site-specific DNA-binding proteins find their targets?
{\it{Nucl. Acid. Res.}} {\bf{32}}:3040-3052.

\bibitem{mirny_review09}
Mirny, L., M. Slutsky, Z. Wunderlich, A. Tafvizi, ..., A. Kosmrlj. 2009.
How a protein searches for its site on DNA: the mechanism of facilitated diffusion.
{\it{J. Phys. A.}} {\bf{42}}:434013.

\bibitem{mirny04}
Slutsky, M., and L.~A. Mirny. 2004.
Kinetics of Protein-DNA Interaction: Facilitated Target Location in Sequence-Dependent Potential.
{\it{Biophys. J.}} {\bf{87}}:4021-4035.

\bibitem{mirny04a}
Slutsky, M., M. Kardar, and L.~A. Mirny. 2004.
Diffusion in correlated random potentials, with applications to DNA.
{\it{Phys. Rev. E}}. {\bf{69}}:061903.

\bibitem{mirny10}
Mirny, L.~A. 2010.
Nucleosome-mediated cooperativity between transcription factors.
{\it{Proc. Natl. Acad. Sci. U.S.A.}} {\bf{107}}:22534-22539.

\bibitem{tolya08}
Cherstvy, A.~G., A.~B. Kolomeisky, and A.~A. Kornyshev. 2008.
Protein-DNA Interactions:  Reaching and Recognizing the Targets.
{\it{J. Phys. Chem. B}}. {\bf{112}}:4741-4750. 

\bibitem{tolya10}
Das, R.~K., and A.~B. Kolomeisky. 2010.
Facilitated search of proteins on DNA: correlations are important.
{\it{Phys. Chem. Chem. Phys.}} {\bf{12}}:2999-3004.

\bibitem{grosberg06}
Hu, T., A.~Yu. Grosberg, and B.~I. Shklovskii. 2006.
How Proteins Search for Their Specific Sites on DNA: The Role of DNA Conformation.
{\it{Biophys. J.}} {\bf{90}}:2731-2744.

\bibitem{kafri09}
Sheinman, M., and Y. Kafri. 2009. 
The effects of intersegmental transfers on target location by proteins.
{\it{Phys. Biol.}} {\bf{6}}:016003.

\bibitem{cox06} 
Wang, Y.~M., R.~H. Austin, and E.~C. Cox. 2006.
Single Molecule Measurements of Repressor Protein 1D Diffusion on DNA.
{\it{Phys. Rev. Lett.}} {\bf{97}}:048302. 

\bibitem{xie09} 
Blainey, P.~C., G. Luo, S.~C. Kou, W.~F. Mangel, ..., X.~S. Xie. 2009. 
Nonspecifically bound proteins spin while diffusing along DNA. 
{\it{Nature Struct. Mol. Biol.}} {\bf{16}}:1224-1229. 

\bibitem{xie07}
Elf, J., G.-W. Li, and X.~S. Xie. 2007.
Probing Transcription Factor Dynamics at the Single-Molecule Level in a Living Cell.
{\it{Science}}. {\bf{316}}:1191-1194.
 
 \bibitem{mirny08}
 Tafvizi, A., F. Huang, J.~S. Leith, A.~R. Fersht, ..., A.~M. van Oijen. 2008. 
 Tumor Suppressor p53 Slides on DNA with Low Friction and High Stability.
 {\it{Biophys. J.}} {\bf{95}}:L01-L03.

\bibitem{zhuang08}
Liu, S., E.~A. Abbondanzieri, J.~W. Rausch, S.~F.~J. Le Grice, and X. Zhuang. 2008. 
Slide into Action: Dynamic Shuttling of HIV Reverse Transcriptase on Nucleic Acid Substrates.
{\it{Science}}. {\bf{322}}:1092-1097.

\bibitem{plischke}
M. Plischke and B. Bergersen, {\it{Equilibrium Statistical Physics}}
(World Scientific, Singapore, 1994). 

\bibitem{misha}
Lukatsky, D. B., and M. Elkin. Energy fluctuations shape free energy of biomolecular interactions. 
arXiv:1101.4529v1. 

\bibitem{lee07}
Lee, W.,  D. Tillo, N. Bray, R. H. Morse, R. W. Davis, ..., C. Nislow. 2007.
A high-resolution atlas of nucleosome occupancy in yeast.
{\it{Nature Gen.}} {\bf{39}}:1235-1244.

\bibitem{struhl95}
Iyer, V., and K. Struhl. 1995. 
Poly(dA:dT), a ubiquitous promoter element that stimulates transcription via its intrinsic DNA structure.
{\it{EMBO J.}} {\bf{14}}:2570-2579.


\end{thebibliography}
\end{document}